\documentclass[twocolumn]{aastex62}
\pdfoutput=1 
\usepackage{amsmath,amstext}
\usepackage[T1]{fontenc}
\usepackage{apjfonts} 
\usepackage[figure,figure*]{hypcap}
\usepackage{graphicx}
\usepackage{color}

\def\gtaprx {\lower .1ex\hbox{\rlap{\raise .6ex\hbox{\hskip .3ex
				{\ifmmode{\scriptscriptstyle >}\else
					{$\scriptscriptstyle >$}\fi}}}
		\kern -.4ex{\ifmmode{\scriptscriptstyle \sim}\else
			{$\scriptscriptstyle\sim$}\fi}}}
\def\ltaprx {\lower .1ex\hbox{\rlap{\raise .6ex\hbox{\hskip .3ex
				{\ifmmode{\scriptscriptstyle <}\else
					{$\scriptscriptstyle <$}\fi}}}
		\kern -.4ex{\ifmmode{\scriptscriptstyle \sim}\else
			{$\scriptscriptstyle\sim$}\fi}}}
\newcommand{\lSect}[1]{{\label{sec:#1}}}

\newcommand{\CASTRO}{\texttt{CASTRO}}
\newcommand{\KEPLER}{\texttt{KEPLER}}

\newcommand{\Ni}{{\ensuremath{^{56}\mathrm{Ni}}}}
\newcommand{\Fe}{{\ensuremath{^{56}\mathrm{Fe}}}}

\newcommand{\Co}{{\ensuremath{^{56}\mathrm{Co}}}}
\newcommand{\He}{{\ensuremath{^{4} \mathrm{He}}}}
\newcommand{\Hy}{{\ensuremath{^{1} \mathrm{H}}}}
\newcommand{\Ox}{{\ensuremath{^{16}\mathrm{O}}}}
\newcommand{\Cr}{{\ensuremath{^{48}\mathrm{Cr}}}}
\newcommand{\Ti}{{\ensuremath{^{44}\mathrm{Ti}}}}
\newcommand{\Si}{{\ensuremath{^{28}\mathrm{Si}}}}
\newcommand{\Ca}{{\ensuremath{^{40}\mathrm{Ca}}}}
\newcommand{\Mg}{{\ensuremath{^{24}\mathrm{Mg}}}}
\newcommand{\Cx}{{\ensuremath{^{12}\mathrm{C}}}}

\newcommand{\erg}{{\ensuremath{\mathrm{erg}}}}

\newcommand{\gcc}{\ensuremath{\mathrm{g}\,\mathrm{cm}^{-3}}}
\newcommand{\kms}{\ensuremath{\mathrm{km}\,\mathrm{s}^{-1}}}

\newcommand{\Msun}{\ensuremath{M_\odot}}

\newcommand{\note}[1]{\emph{\textcolor{red}{}}}
\newcommand{\cw}[1]{\emph{\textcolor{blue}{}}}

\makeatletter

\newcommand{\Rmnum}[1]{\expandafter\@slowromancap\romannumeral #1@}
\newcommand{\ken}[1]{\textcolor{black}{ #1}}
\newcommand{\djw}[1]{\textcolor{black}{#1}}
\newcommand{\dw}[1]{\textcolor{black}{#1}}

\makeatother

\shorttitle{3D Magnetar-Powered Supernovae}
\shortauthors{Chen, Woosley \& Whalen}

\begin{document}

\title{Three-Dimensional Simulations of Magnetar-Powered Superluminous  Supernovae}

\author{Ke-Jung Chen}
\affiliation{Institute of Astronomy and Astrophysics, Academia Sinica,  Taipei 10617, Taiwan}
\affiliation{Division of Theoretical Astronomy, 
	National Astronomical Observatory of Japan, 
	Tokyo 181-8588, Japan} 
\author{S. E. Woosley}
\affiliation{Department of Astronomy and Astrophysics,
	University of California at Santa Cruz, Santa Cruz, CA 95060, USA}

\author{Daniel J. Whalen}
\affiliation{Institute of Cosmology and Gravitation, University of Portsmouth, Portsmouth PO1 
3FX, UK}
\affiliation{Ida Pfeifer Professor, Department of Astrophysics, University of Vienna, Tuerkenschanzstrasse 17, 1180, Vienna, Austria}

\correspondingauthor{Ke-Jung Chen}
\email{kjchen@asiaa.sinica.edu.tw}

\begin{abstract}

A rapidly spinning magnetar in a young supernova (SN) can produce a superluminous transient by converting a fraction of its rotational energy into radiation.  Here, we present the first three-dimensional hydrodynamical simulations ever performed of a magnetar-powered SN in the circumstellar medium formed by the ejection of the outer layers of the star prior to the blast.  We find that hydrodynamical instabilities form on two scales in the ejecta, not just one as in ordinary core-collapse SNe:  in the hot bubble energized by the magnetar and in the forward shock of the SN as it plows up ambient gas.  Pressure from the bubble also makes the instabilities behind the forward shock more violent and causes more mixing in the explosion  than in normal SNe, with important consequences for the light curves and spectra of the event that cannot be captured by one-dimensional models.  We also find that the magnetar can accelerate Ca and Si to velocities of $\sim $ 12000 $\kms$ and account for their broadened emission lines in observations.  Our simulations also reveal that energy from even weak magnetars can accelerate iron-group elements deep in the ejecta to $5000-7000$ \kms\ and explain the high-velocity Fe observed at early times in some core-collapse SNe such as SN 1987A.    

\end{abstract}


\section{Introduction}

Magnetars are neutron stars (NSs) with magnetic fields exceeding 10$^{13}$ Gauss (G).  Their unusually strong fields may come from the collapse of a rapidly, differentially-rotating iron core \citep{Dun92,Tho93,Whe00,Tho04} during a supernova explosion \citep{Kou98}.  Magnetars are therefore often born with high rotation rates and in some extreme cases can have magnetic fields $\gtrsim 10^{16}$ G and periods of $\lesssim 1$ ms.   At spin-down rates of $\sim 10^{51}$ erg s$^{-1}$ these 'millesecond magnetars' can lose most  of their energy in $\sim$ 20 s and may be responsible for some long-duration gamma-ray bursts \citep[e.g.,][]{Met11,Met15}.  However, most magnetars have magnetic fields of $10^{14}-10^{15}$ G and periods of $\sim 1-10$ ms. If a magnetar radiates away its rotational energy in the same way as a pulsar, its luminosity lasts for much longer times in lower B-fields.  

\citet{Mae07}, \citet{Woo10} and \citet{Kas10} have shown that the birth of a magnetar can power exceptionally luminous transients \citep[superluminous supernovae, or SLSNe;][]{Qui11,Gal12,Ins13} because a significant fraction of the magnetar's spin energy can be emitted as optical radiation at a later time.  Light curves from one-dimensional (1D) magnetar models by  \citet{Kas16}, \citet{Mor17} and \citet{Des18, Des19} are in general agreement with recent SLSN observations by \citet{Ins16}, \citet{Ins17}, \citet{Mar18}, \citet{Nic19} and \citet{Mar19}. 

However, there are important limitations to 1D light curve models and their underlying hydrodynamical solutions.  The magnetar deposits its spin-down energy in a small volume, and when the energy of this hot bubble becomes comparable to the kinetic energy of the surrounding ejecta it begins to plow it up.  In 1D models the ejecta that accumulates on the surface of the bubble collapses into a thin shell with overdensities $\Delta_S = <\delta \rho/\rho>$ that can exceed 1000. If such densities were achieved in real explosions most of the radiation would be trapped by the shell and the SN spectra would indicate that most of the ejecta is moving at a single speed, which is not supported by observations.  Furthermore, it is difficult to explain the light curves of some superluminal events with 1D magnetar models.  The recent discovery of the unusual SN IPTF14hls \citep{14hls}, whose light curve lasts for long times and exhibits multiple peaks, is one such case \citep{Che18, Mor18}. Although the peaks of this light curve may be due to episodic bursts from a central magnetar or multiple collisions of ejecta with circumstellar structures \citep[e.g.]{woosley2007,chen2014b, Woo18}, they may also be due to inhomogeneous emission from clumpy ejecta if the shell breaks up.  

Past studies have clearly shown that the dynamics of magnetar-driven explosions can only be captured by multidimensional simulations.  \citet{Che82}, \citet{Che92}, \citet{Jun98}, \citet{Blo01} and \citet{Che11} found that collisions of pulsar winds with circumstellar media are subject to hydrodynamic instabilities.  More recent studies have examined magnetar winds embedded in young SN remnants with multidimensional simulations \citep{Che16, Che17,Kas16, Blo17, Suz17, Suz18}.  \citet{Che16} considered a magnetar with a 4 $\times$ 10$^{14}$ G field and initial spin period of 1 and 5 ms.  They found that mixing in this explosion is strong enough to fracture the accelerating shell into filamentary structures.  However, with the exception of some local 3D simulations \citep{Blo17} these studies were all limited to 2D and short times \citep{Che16}.  \dw{To understand the spectra of these explosions they must be modeled in 3D over long times to (1) capture the entirety of their light curves and (2) properly simulate their hydrodynamical instabilities, which govern their light curves and are very different in 3D than in 2D.}  
  
We have performed new high-resolution simulations of a magnetar-powered SN, evolving it for $\sim$ 200 days after NS formation to follow the core-collapse (CC) SN shock and the matter accelerated by the magnetar wind behind it.  These models are a considerable improvement over \citet{Che16} because they are performed in 3D in a $4\pi$ geometry, not in 2D in just one octant, and they have a domain that is approximately 10$^5$ times larger than the radius of the original star.  This large domain allows us to follow the mixing of ejecta for more than 200 days after magnetar formation, far longer than the few hours in our previous study.  We discuss our progenitor model and numerical methods in Section 2 and examine the evolution of the explosion in Section 3.  We discuss the consequences of our models for the spectra and light curves of magnetar-powered SLSNe and conclude in Section 4.

\section{Numerical Method}
\lSect{num_method}

We simulate the SN in two stages.  The progenitor star is first exploded in the 1D \KEPLER\ stellar evolution code.  A short time after shock breakout from the surface of the star  we turn on luminosity from the magnetar and continue to evolve the explosion until nuclear burning driven by energy from the magnetar is complete.  At this stage we map the 1D blast profiles from \KEPLER\ onto a 3D nested grid in the \CASTRO\ code and evolve it out to 200 days, continuing the injection of energy by the magnetar begun in \KEPLER.  Turning on the magnetar in \KEPLER\ until all nuclear burning is complete relieves us of the cost of solving a reaction network in 3D in \CASTRO\ without any loss in accuracy because nucleosynthesis driven by radiation from the magnetar is complete well before any fluid instabilities can form in the thin shell plowed up by this energy.

\subsection{Magnetar Luminosity}
\lSect{mag}

We use the Larmor formula for the luminosity of the magnetar \citep{Lyn90,Che16}:
\begin{equation}
\begin{split}
\rm L_m & = \frac{32\pi^4}{3c^2} \rm (BR_{ns}^3\sin\alpha)^2P^{-4}
\\ & \rm \sim 1.0 \times 10^{49}B_{15}^2 P_{ms}^{-4} \quad \erg\ s^{-1},
\end{split}
\label{eq:magP}
\end{equation}
where $B_{15}$ is the surface dipole field strength at the equator of the NS in units of 10$^{15}$ G, $P_{\rm ms}$ is the initial spin period in milliseconds, $R_{\rm ns}$ is the radius of the NS and $\alpha$ is the inclination angle between the spin and magnetic axes.  In our models, $R_{\rm ns} = 10^{6}$ cm and $\alpha = 30^{\circ}$.  The spin energy is
\begin{equation}
\rm E = \frac{1}{2}I\omega^2 \sim 2\times 10^{52}P_{ms}^{-2} \quad
\erg.
\label{eq:magE}
\end{equation}
If the magnetic field is constant then the evolution of the luminosity, period, and spin energy of the magnetar over time can be approximated by 
\begin{equation}
\begin{split}
L_{\rm M}(t) & \sim (1+t/t_m)^{-2}E_0t_m^{-1}\ \rm erg\ s^{-1},\\
P(t) & \sim (1+t/t_m)^{1/2}P_0\  \rm ms,\\ 
E(t) & \sim (1+t/t_m)^{-1}E_0\ \rm erg,
\end{split}
\label{eq:evo}
\end{equation}
where $P_0 = P_{\rm ms}(0)$, $E_0=E(P_0)$ and $t_{\rm m} \sim 2 \times 10^3 P_{\rm ms}^2B_{15}^{-2}$ is the magnetar spin-down timescale. 

\begin{figure*}
\centering
\includegraphics[width=\columnwidth]{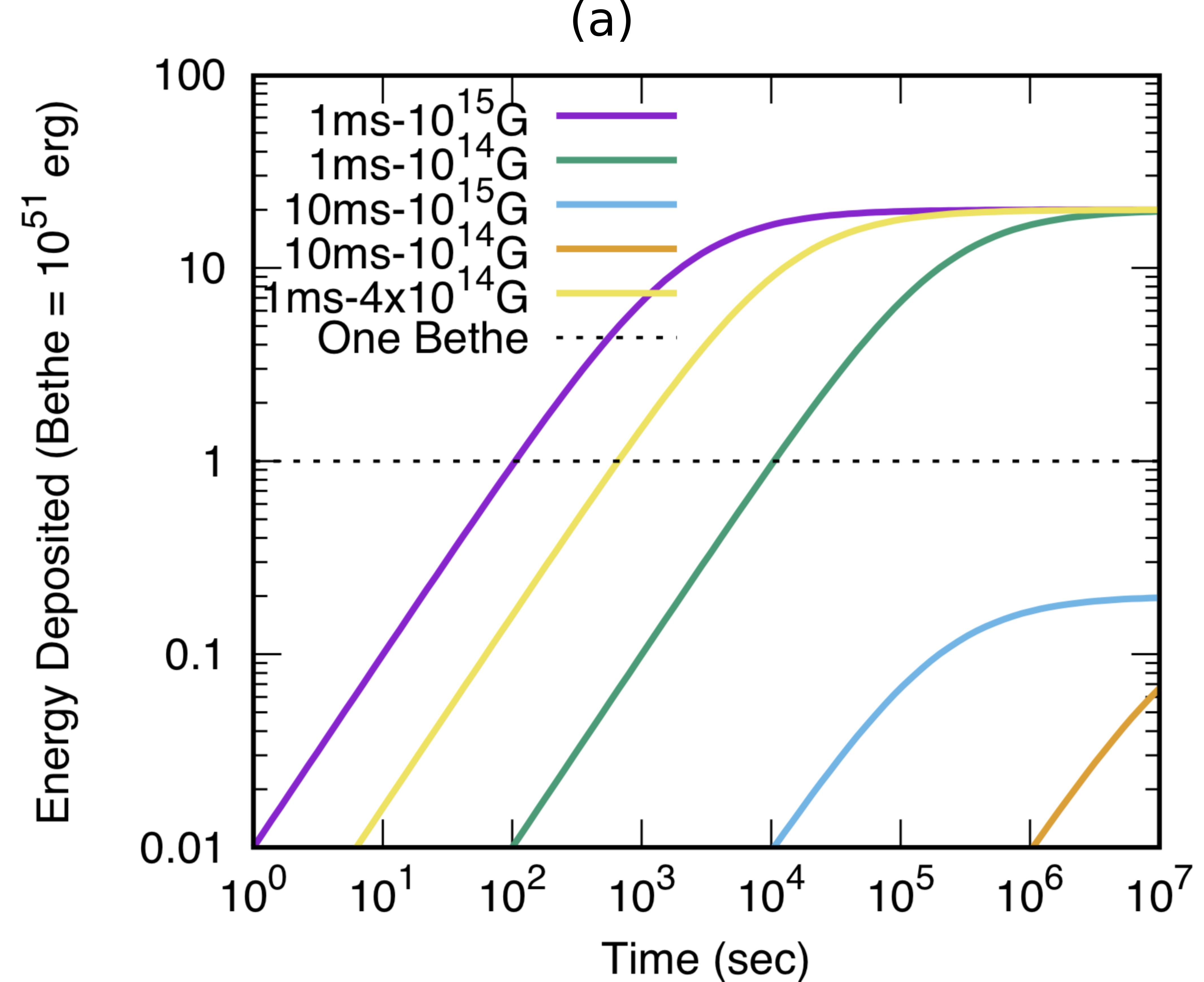}	
\includegraphics[width=\columnwidth]{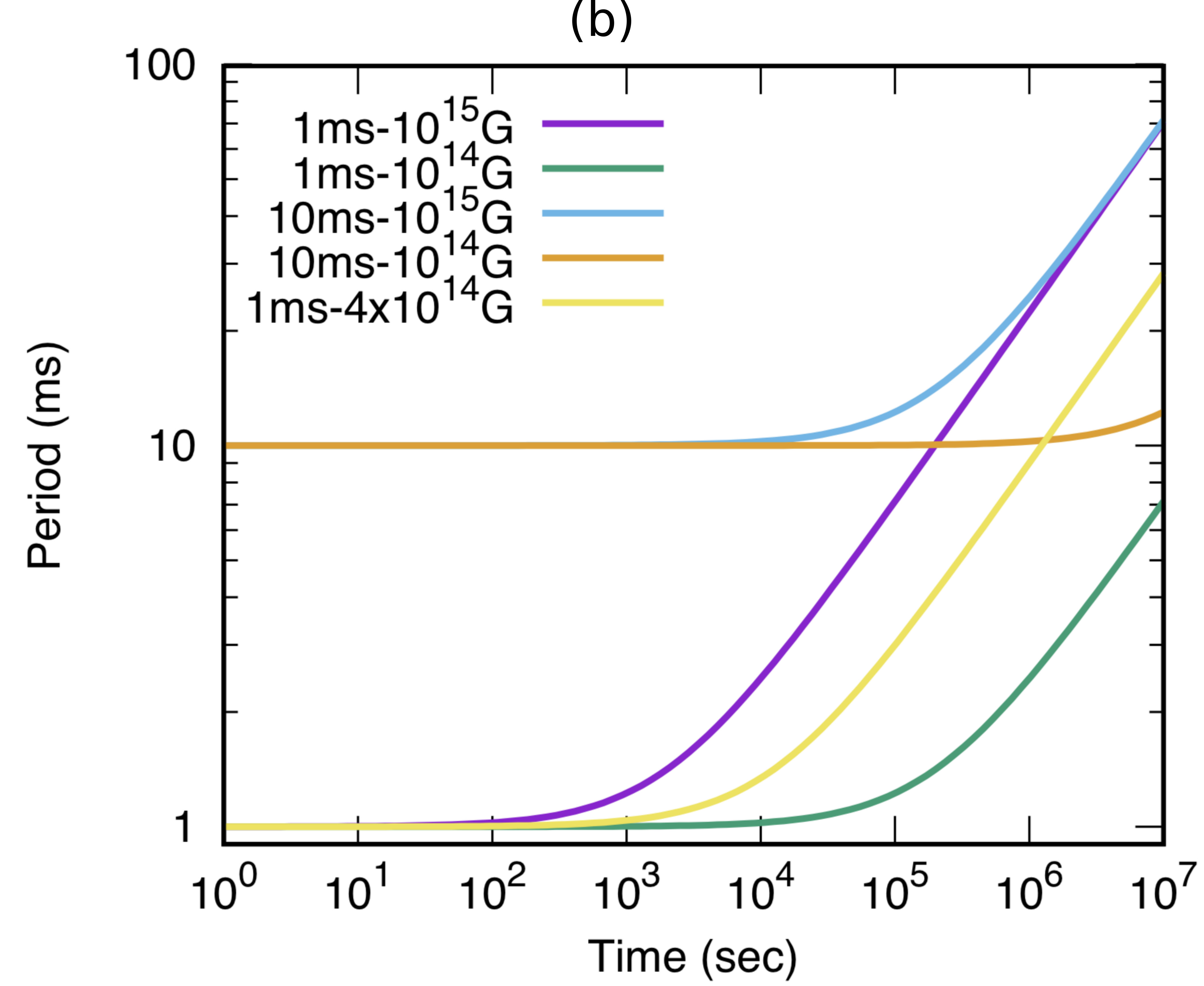}
\includegraphics[width=\columnwidth]{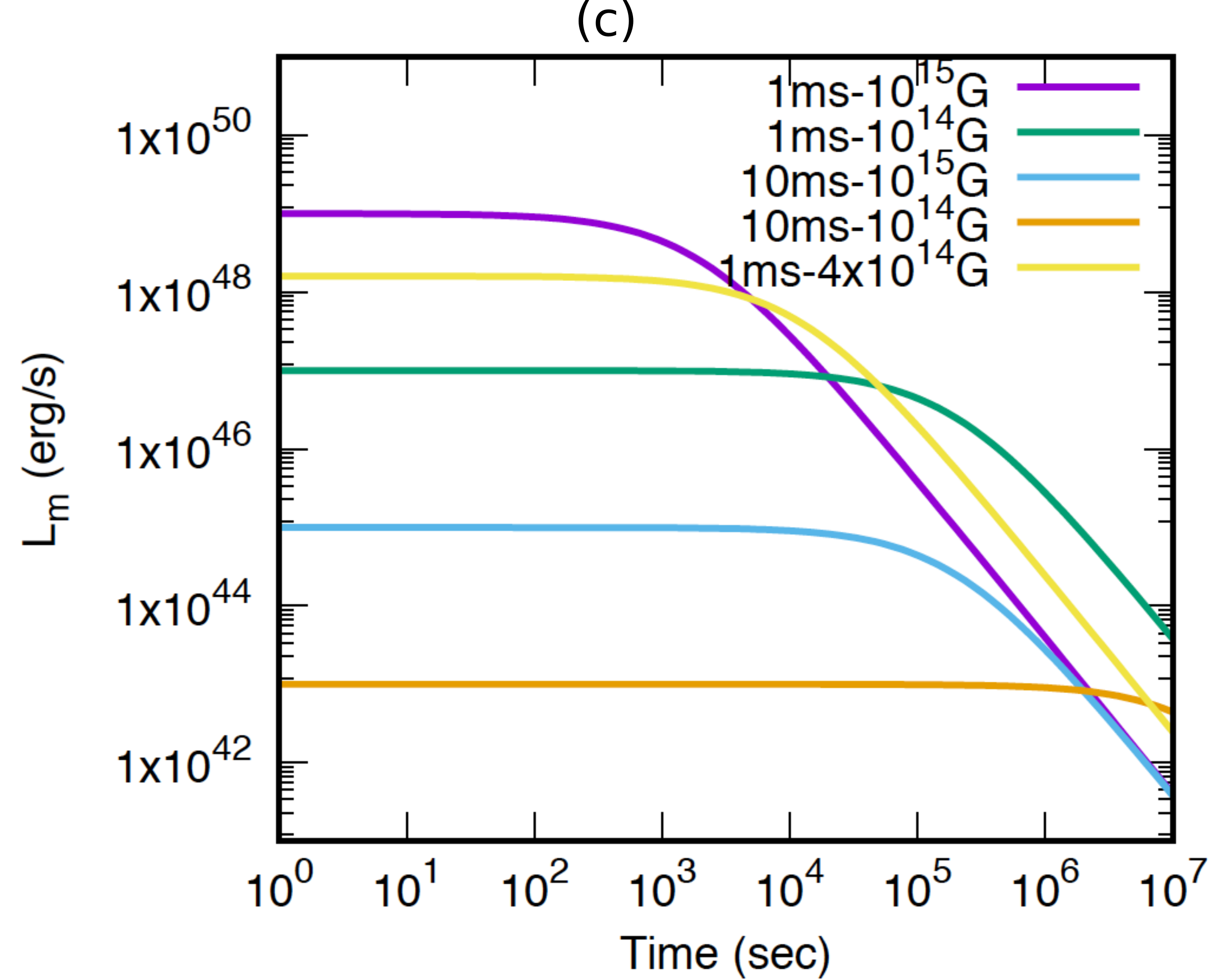} 
\includegraphics[width=\columnwidth]{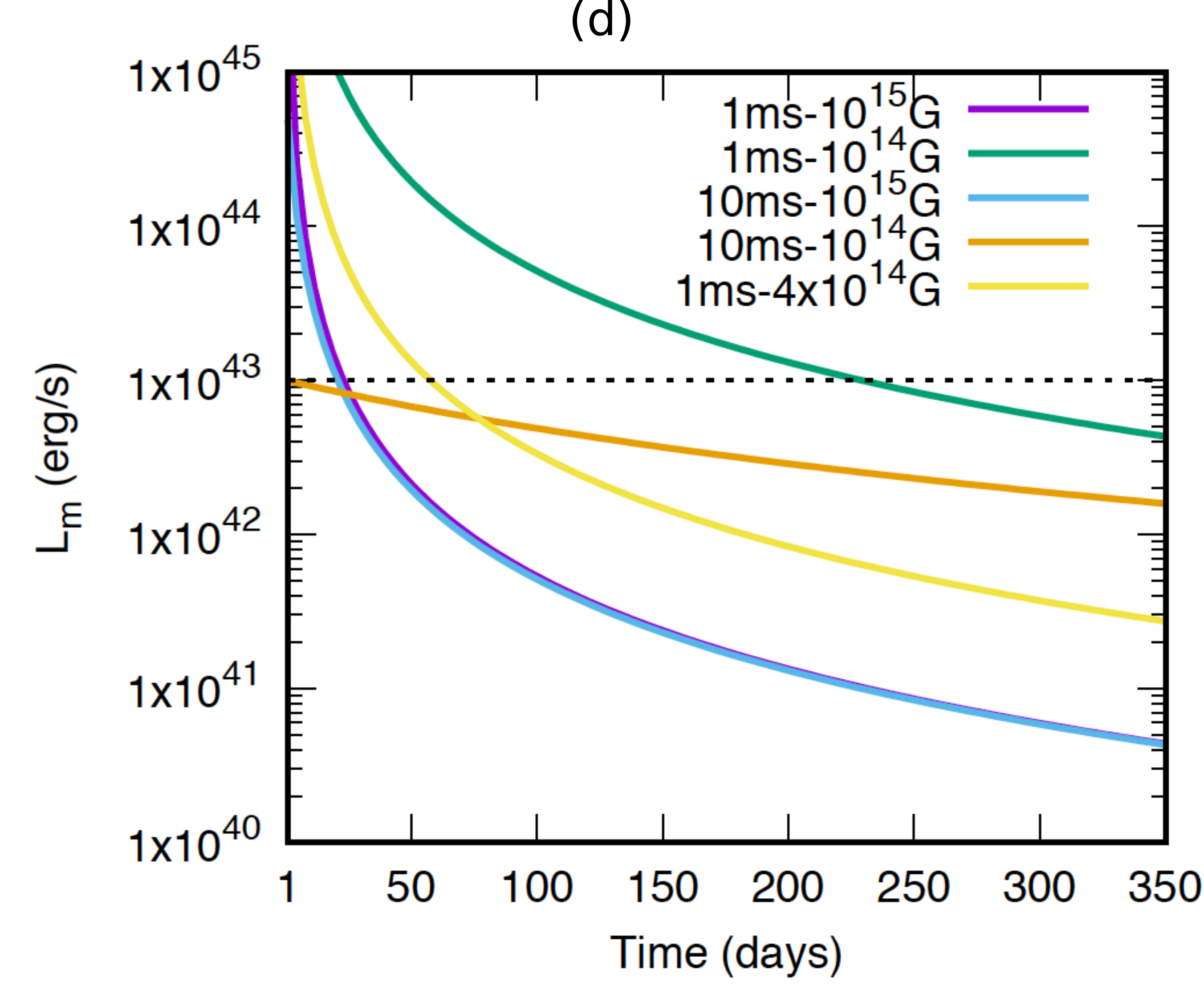}
\caption{Evolution of the dipole radiation from a magnetar for a variety of spin rates and magnetic fields: (a) total energy deposited, (b) period, (c) luminosity, (d) same as (c) but on a linear scale in time. }
\label{fig:dipole}
\end{figure*}

The evolution of the magnetar is shown in Figure~\ref{fig:dipole} for several magnetic fields and initial spins.  The spin energies of the 1 and 10 ms magnetars are $\sim$ $2 \times 10^{52}$ and $2\times10^{50}$ erg, respectively.  Their initial luminosities can vary from $\sim 10^{42} - 10^{49}$ erg s$^{-1}$ depending on period and magnetic field.  The 1-ms magnetar with a magnetic field of $10^{14} - 10^{15}$ G can release $10^{51}$ erg in $10^2 - 10^4$ s, with the remainder being emitted in just $10^3 - 10^7$ s.  Consequently, dipole radiation from the magnetar can efficiently convert its spin energy to luminosity.  

Which choice of magnetar parameters is most likely to produce a superluminous event?  We show the magnetar luminosity over time in days in Figure~\ref{fig:dipole}c.  The average luminosity of the SN itself at early times is plotted as the horizontal dashed line at $10^{43}$ erg s$^{-1}$.  The magnetar produces a superluminous event only if its luminosity exceeds that of the SN by the time photons from the NS can diffuse out of the ejecta, typically on timescales of 50 - 100 days.  How much of this energy is converted into kinetic energy of the ejecta or escapes as SN luminosity in any given event remains highly uncertain.  However, if, in the most optimistic scenario, all of the energy of the magnetar is emitted as radiation only the $1 - 4 \times 10^{14}$ G, 1 ms magnetars are brighter than the SN after 50 days.  Those with higher fields or longer periods lose energy much earlier and are no longer bright by the time their photons exit the ejecta.  We therefore adopt the $4 \times 10^{14}$ G, 1 ms magnetar as the fiducial case in our study, which has a total spin energy of $\sim$ $2 \times 10^{52}$ erg.

Although it is thought that the spin-down energy of the magnetar is converted into high-energy photons, neutrinos, cosmic rays and winds, how it is partitioned among them is not well understood and the spectrum of the photons themselves is not well-constrained by models.  Furthermore, the efficiency with which photons, particles and winds dynamically couple to local ejecta is not completely clear.  Most calculations of magnetar-powered SLSN light curves assume that 100\% of the dipole radiation goes into luminosity when in reality some goes into $PdV$ work on the ejecta that changes its dynamics.  We deposit all the energy of the magnetar into the ejecta as heat for simplicity.  This approximation yields an upper limit to the violence of any fluid instabilities that could affect the light curve of the explosion because none of the energy of the magnetar directly escapes the flow as radiation.

It is not known for certain how much time is required for a proto-NS to evolve into a magnetar so we turn it on in \KEPLER\ $\sim$ 100 s after the explosion, just after the shock has broken out of the surface of the star and the ejecta is expanding nearly homologously.  This choice is reasonable, given that even if the magnetar had formed instantaneously it would have emitted at most $0.6\%$ of its total spin energy in the first 100 s, which is negligible in comparison to the energy it later deposits in the flow or even the much lower energy of the original explosion.  

\subsection{\KEPLER}
\lSect{presn}

We consider the same star as \citet{Che16}, a 6 \Msun\ carbon-oxygen (CO) core that evolved from a 24 \Msun\ zero-age main sequence star that loses its hydrogen envelope and part of its helium envelope prior to explosion \citep{Suk14}.  The core has a radius of $2 \times 10^{11}$ cm and $\Cx$ and $\Ox$ mass fractions of 0.14 and 0.86, respectively. This star evolves through carbon, neon, oxygen, and silicon burning, building up a 1.45 \Msun\ iron core prior to collapse.  We use a CO star because many SLSNe have been found to be Type I, implying that the progenitor has lost its outer layers to a stellar wind or binary interaction.  It was evolved until its iron core reached collapse velocities of $\sim$ 1,000 km s$^{-1}$, at which point the SN was initialized by the injection of linear momentum in its core.  

CC SNe are not emergent features of \KEPLER\ simulations because it lacks the physics and dimensionality required to capture core bounce and the heating of the atmosphere of the proto-NS by neutrinos, both of which are thought to launch the SN shock.  We instead trigger the explosion by injecting linear momentum into the core at a predetermined mass coordinate with an energy corresponding to the energy chosen for the explosion.  Here, the CO star explodes as a core-collapse (CC) SN with an energy of 1.2 $\times$ 10$^{51}$ erg, producing a 1.45 \Msun\ NS (magnetar) and ejecting 0.22 \Msun\ of \Ni.  \KEPLER\ includes the gravity due to the NS and the self-gravity of the ejecta.  

Multidimensional simulations of the evolution of the cores of massive stars have shown that  convective mixing of the oxygen and silicon layers occurs prior to the collapse of the iron core \citep{Ma07}.  The mechanics of the explosion can be sensitive to this convective structure.  Furthermore, numerous studies of core collapse and bounce have found that fluid instabilities and mixing occur even at these early stages of CC SNe \citep{burrows1995,janka1996,mezz1998,woosley2005,kif06}.  Even on short time scales of several ms, accretion shock instabilities along with neutrino heating produce strong mixing and break the spherical symmetry of the core.  \KEPLER\ cannot reproduce any of this mixing in 1D from first principles so we artificially mix some of the \Si\ core using the mixing length theory in \KEPLER. 

\begin{figure}
\centering
\includegraphics[width=\columnwidth]{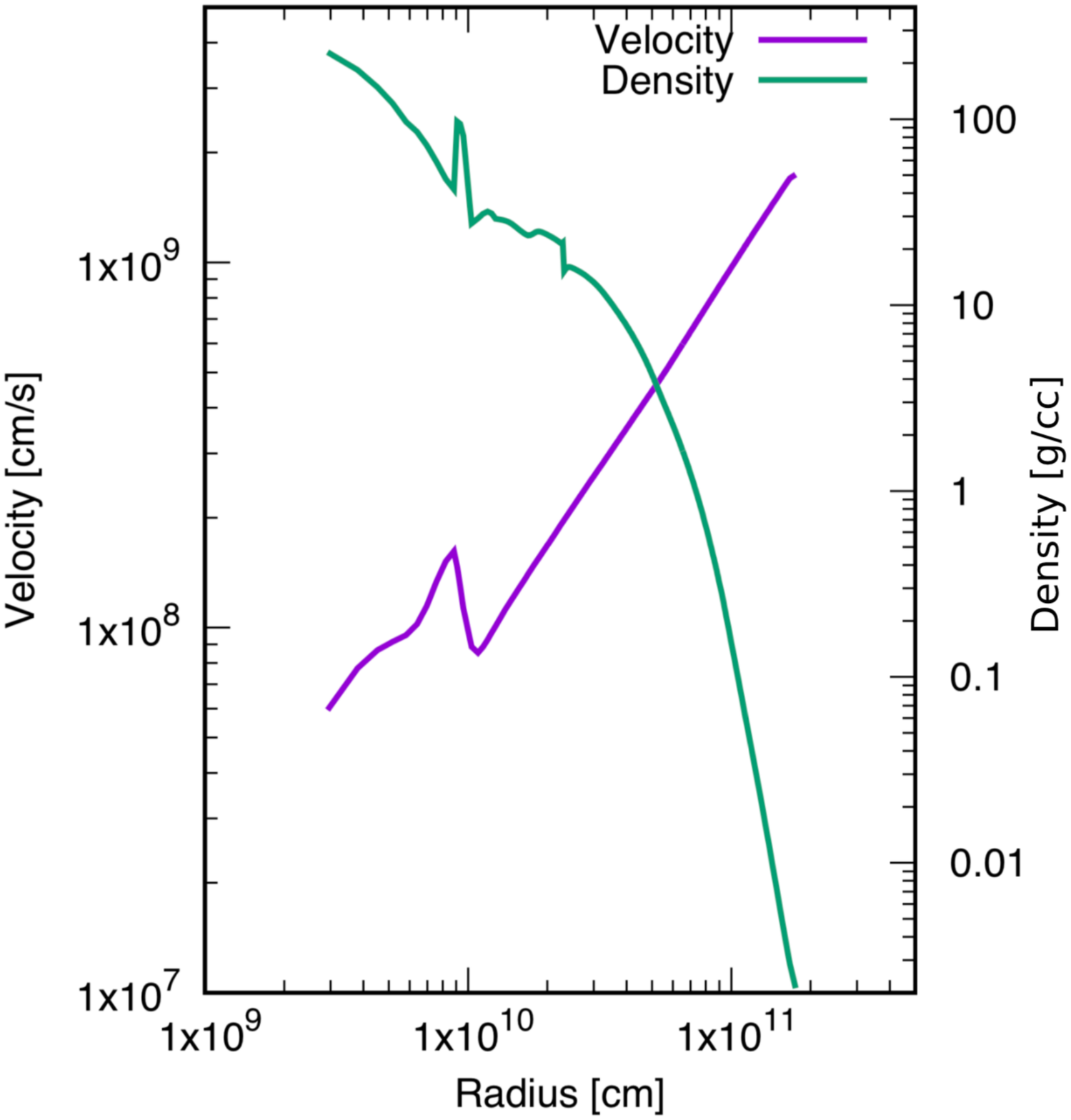} 
\vspace{-0.5in}
\caption{Velocity and density profiles of the magnetar-driven SN in \KEPLER\ 110 seconds after the explosion.  A dense shell driven by luminosity from the magnetar appears at $\sim 10^{10}$ cm approximately 10 seconds after the formation of the magnetar. Gas heated by the magnetar plows up the surrounding medium, forming a shock and creating the velocity jump at $\sim 10^{10}$ cm.}
\label{fig:ic}
\end{figure}

As noted above, the magnetar is turned on 100 s after linear momentum is injected into the core of the star to trigger the SN.  Its luminosity is uniformly deposited as heat in the innermost 10 Lagrangian zones of the mesh, which enclose the central 0.12 \Msun\ of the ejecta.  These zones have an approximately constant energy generation rate per gram.  The total energy deposited in the ejecta evolves over time as shown in the plot in Figure~\ref{fig:dipole}c corresponding to the magnetic field and period of our magnetar.  Density and velocity profiles of the ejecta 110 seconds after the explosion are shown in Figure~\ref{fig:ic}.  

The velocity of the forward shock is $\sim$ $2 \times 10^{9}$ cm s$^{-1}$ and the innermost ejecta is expanding at $\sim 10^{8}$ cm s$^{-1}$.  Density and velocity jumps appear at $\sim$ 10$^{10}$ cm as luminosity from the magnetar heats the inner ejecta and it expands, plowing up gas and forming a dense shell that moves supersonically with respect to the ejecta.  If allowed to evolve further in 1D, the shell would reach $\Delta_S \sim$ 1000 - 2000 within a few thousand seconds after the explosion \citep[see Figure~4 of][]{Che16}.  We instead halt the simulation and port these profiles into \CASTRO\ because all nuclear burning due to energy from the magnetar has ceased but the shell is not yet prone to hydrodynamical instabilities. 

\subsection{\CASTRO}
\lSect{castro}

We map 1D spherically-symmetric \KEPLER\ blast profiles onto 3D cartesian grids in \CASTRO\ with a conservative scheme developed by \citet{Che13} that preserves the mass, momentum and energy of the ejecta.  \CASTRO\ is a multi-dimensional AMR hydrodynamics code \citep{Alm10, Zha11} with an unsplit piecewise-parabolic method (PPM) hydro scheme \citep{Woo84}, multi-species advection, and several equations of state (EOS). At early stages of the simulation we use the Helmholtz EOS, which includes electron and positron pairs of arbitrary degeneracy, ions, and radiation \citep{Tim00}, to model the partly degenerate CO core but later switch to an ideal gas EOS after the density of the expanding ejecta falls below 10$^{-12}$ \gcc.  We evolve mass fractions for \Hy, \He, \Cx, \Ox, \Mg, \Si, \Ca, \Ti, \Cr, and \Ni\ to study how they are mixed during the explosion but do not employ a reaction network because nuclear burning due to heating by the magnetar has ceased.  We neglect heating due to radioactive decay of \Ni\ (and its conversion into iron) because the total energy released in our models is less than 1\% of that of the SN.  \ken{The gravities due to the ejecta, its envelope and the compact object are all included in our models.  The self-gravity of the ejecta is calculated from the monopole approximation by constructing a spherically-symmetric gravitational potential from the radial average of the density and then determining the corresponding gravitational force everywhere in the AMR hierarchy.  This approximation is reasonable, given the globally spherical structure of the supernova ejecta.  The magnetar is treated as a point source.}

\begin{figure*}
\centering
\includegraphics[width=\textwidth]{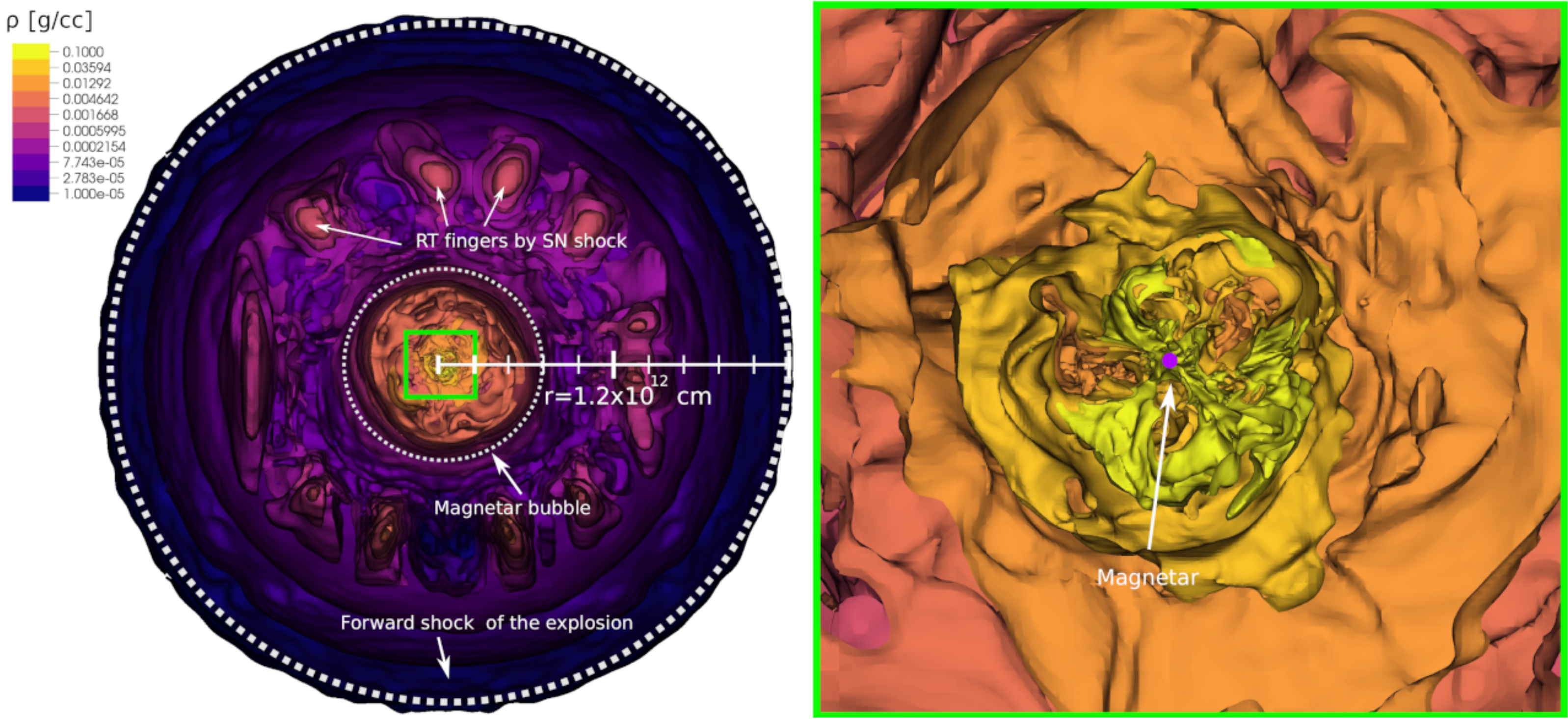} 
\caption{\label{fig:early} 720 seconds after the birth of the magnetar. In the left panel the inner and outer dotted circles mark the radii of the magnetar bubble ($3.6 \times 10^{11}$ cm) and forward shock ($1.2 \times 10^{12}$ cm), respectively.  RT instabilities have appeared behind the forward shock and the interior of the bubble is highly turbulent.  Right panel:  close-up of the bubble, in which turbulent eddies have signficantly mixed the ejecta. \vspace{0.3in}}
\end{figure*}

\begin{figure}
\centering
\includegraphics[width=\columnwidth]{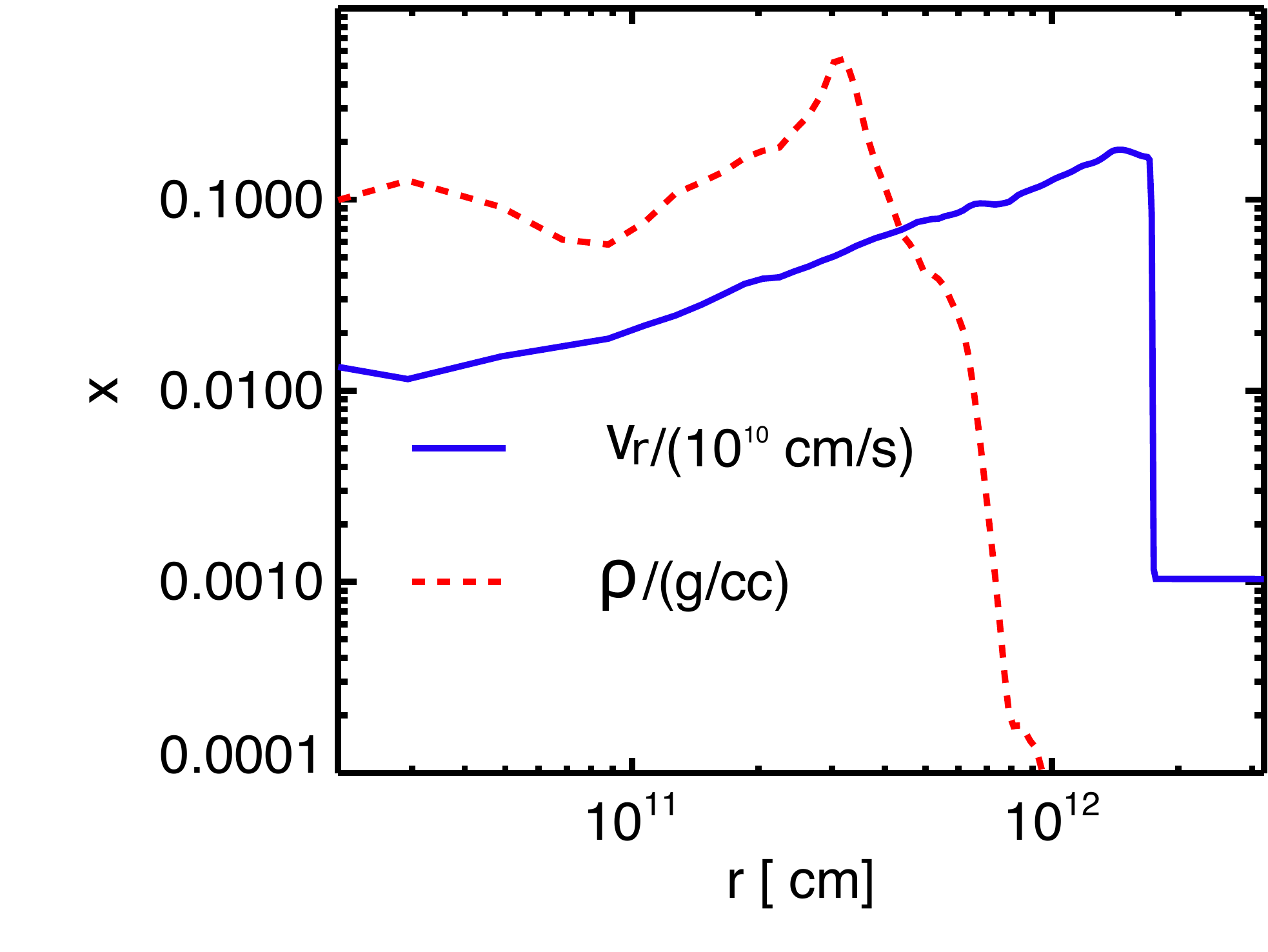} 
\caption{\label{fig:1D_avg} 1D angle-averaged profiles of the density and velocity 720 seconds after the formation of the magnetar.  The density peak at $r \sim 3 \times 10^{11}$ cm is the shell created by the magnetar bubble.}
\end{figure}

We add a circumstellar medium (CSM) to the outer boundary of the ejecta due to the loss of the H and He layers of the star prior to the explosion.  The envelope can be ejected by gravity waves \citep{Qs12, Sq14} or super-Eddington stellar winds \citep{Owo17, Owo19} before the death of the star.  This measure also prevents superluminal velocities when the shock crashes out of the star into very diffuse densities because \CASTRO\ lacks relativistic hydrodynamics.  \cite{Sq14} found that such ejections can create wind profiles with total masses of 10$^{-3}$ - 1 \Msun\ extending out to $10^{15} - 10^{16}$ cm.  We therefore surround the CO core with a $\rho_r =\rho_0(r/r_0)^{-2}$ wind profile, where $\rho_0$ is the density at the outer boundary of the ejecta, which we set to $2.11 \times 10^{-11}$ \gcc\ at $1.75 \times 10^{11}$ cm. $\rho_0$ is a factor of 10$^{-4}$ lower than the density at the surface of the star and represents the abrupt drop in density between the star and the CSM.  The total mass of the envelope is 0.81 \Msun.  

Our \CASTRO\ root grid is $5 \times 10^{12}$ cm on a side with a resolution of 128$^3$ and  eight nested grids centered on the star, which has a radius of $1.75 \times 10^{11}$ cm.  This hierarchy of grids has a maximum initial spatial resolution of $1.53 \times 10^8$ cm to 
resolve the ejecta in which energy from the magnetar is deposited and fluid instabilities later form.  After the launch of the simulation we allow up to 4 additional levels of refinement, which are triggered by gradients in density, velocity and pressure.  \djw{If the SN shock reaches the grid boundary we double the size of the mesh while holding the number of mesh points constant and conservatively map the flows onto this new grid using the approach by \citet{Che13}.  In our run the original box was doubled twelve times to a final size of $2.05 \times10^{16}$ cm.}  Our maximum spatial resolution is similar to the finest zone size in \citet{Che16}, $\sim 1.2\times10^8$ cm, so it is sufficient to capture fluid instabilities and mixing.  Outflow boundary conditions were set on the grid.  The simulations are evolved until the shock reaches $\sim$ 10$^{16}$ cm about 200 days after the explosion.  

The magnetar luminosity that was turned on in \KEPLER\ is continued in \CASTRO\ by depositing heat in a central sphere that has a radius of $\sim 2\%$ that of the expanding ejecta. This sphere is always resolved by $\sim 100$ zones.  We sinusoidally vary the energy we deposit in the sphere in angle by an amplitude of 1\% to approximate anisotropies in emission from the NS.  If densities in the cavity formed by the expansion of the newly-heated ejecta become too small, energy deposition can lead to superluminal velocities because \CASTRO\ is not a relativistic code.  To prevent this we add small amounts of mass with the energy that total $\sim 2.92 \times 10^{-10}$ \Msun\ by the end of the run, which is negligible compared with the mass of ejecta.  This mass injection also limits the maximum velocity of the wind to $\sim 30\%$ speed of the light. 

\section{3D Explosion Dynamics}

We now discuss instabilities and mixing driven by energy from the magnetar at early and late times.

\subsection{Bubble Formation}

The sphere into which the magnetar initially deposits its energy has a radius of $3 \times 10^9$ cm.  At a luminosity of $\sim 10^{48}$ erg s$^{-1}$ it deposits more than $1 \times 10^{51}$ erg  into the flows in the first 1000 s.  This rapid injection of energy causes the ejecta around the magnetar to quickly expand, forming a hot bubble.  Figure~\ref{fig:early} shows densities in the bubble and surrounding ejecta 720 s after the formation of the magnetar.  The radii of the bubble and forward shock marked by the white dashed circles are found by spherically averaging the densities and velocity flows and identifying the appropriate peaks, as in Figure~\ref{fig:ic}.  The forward shock of the SN has grown to $\sim 10^{12}$ cm.  \ken{Rayleigh-Taylor (RT) fluid instabilities have formed on two spatial scales, just behind the forward shock and at the boundary of the magnetar bubble.  The first are driven by the forward shock and the second are driven by wind from the magnetar. Only the former are still visible because the latter have devolved into turbulent hot gas in the bubble, as seen in the eddy-like structures within the inner dashed circle of Figure~\ref{fig:early}.  Both instabilities efficiently mix the ejecta.}

We plot 1D angle-averaged profiles of density and velocity in Figure~\ref{fig:1D_avg}.  The shell driven outwards by the magnetar bubble is visible as the density bump at $r \sim 3\times10^{11}$ cm, and it moves at a velocity of $5 \times 10^8$ cm s$^{-1}$.  The anisotropies in the flows driven by luminosity from the magnetar at early times seen in the right panel of Figure~\ref{fig:early} lead to elongation in the bubble at much later times, as we discuss below.
  
\begin{figure*}
\centering
\includegraphics[width=\columnwidth]{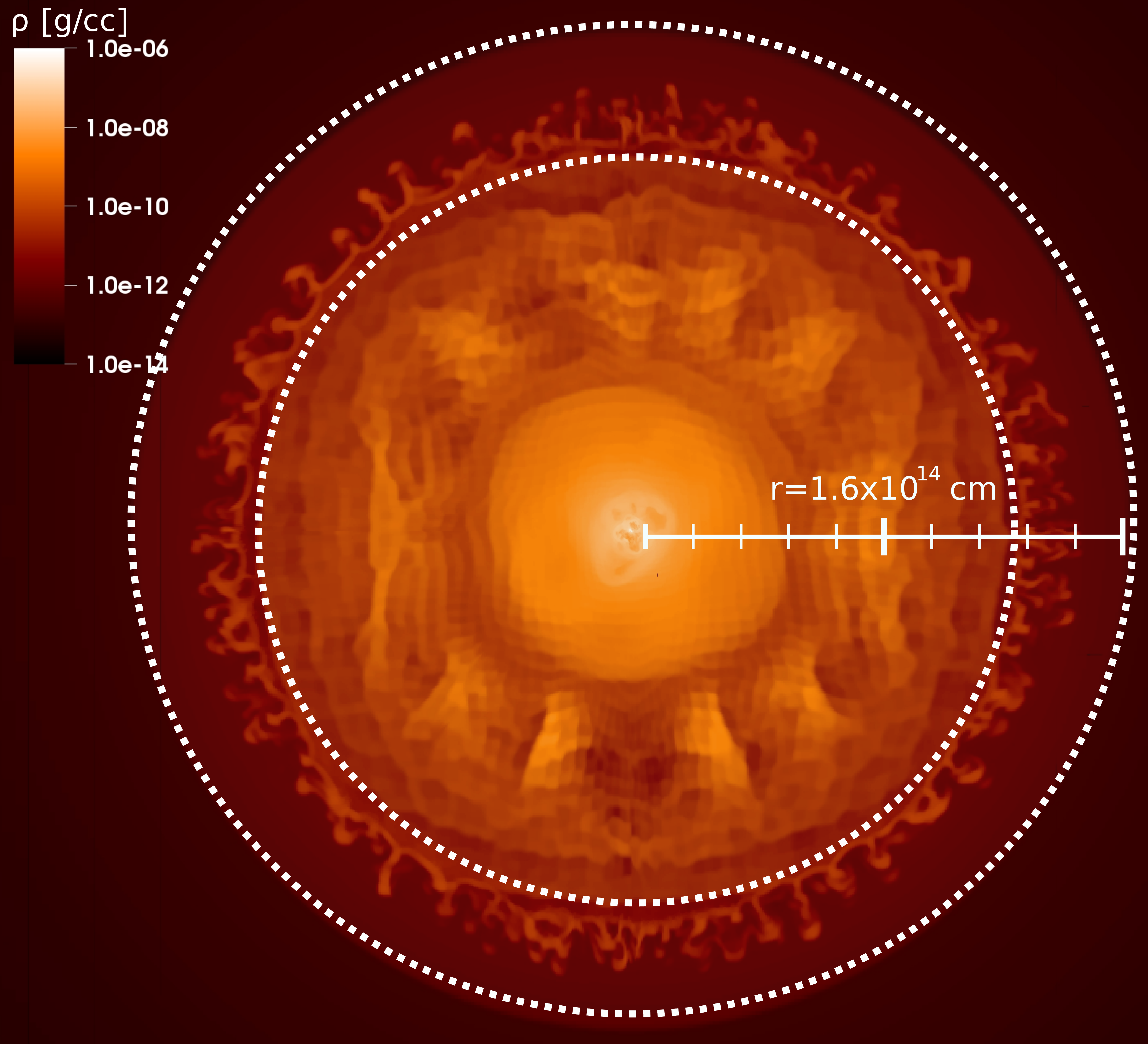} 
\includegraphics[width=.96\columnwidth]{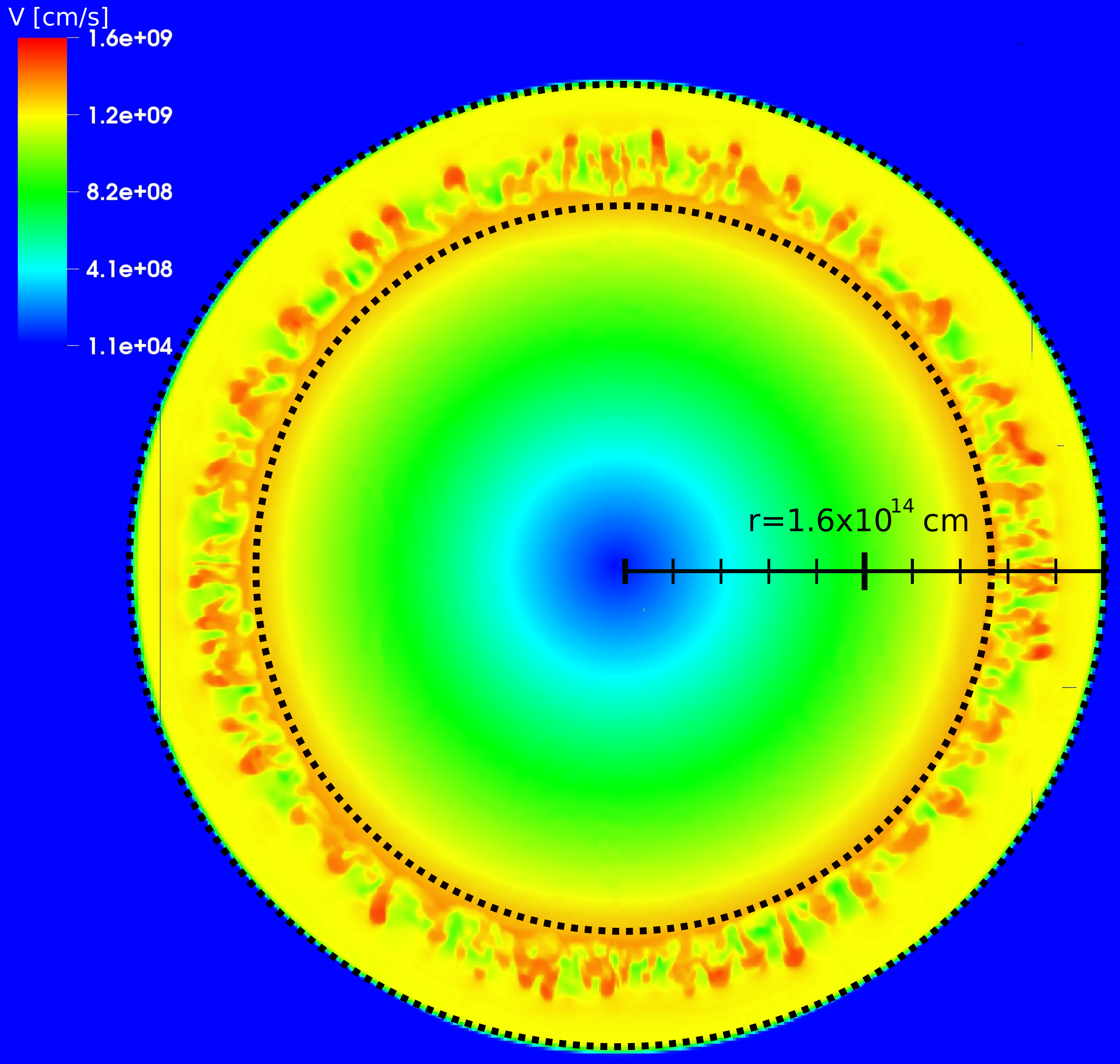} 
\caption{\label{fig:int} \ken{30 hours after magnetar formation. Left panel: densities.  The inner and outer dotted circles mark the radii of accelerated ejecta by the  magnetar wind ($1.25 \times 10^{14}$ cm) and forward shock of the SN ($1.6 \times 10^{14}$ cm).  The fingers at $1.3 - 1.6 \times 10^{14}$ cm are new RT instabilities due to the deceleration of forward shock as it snowplows the CSM.  The more diffuse structures at 0.5 - 1.1 $\times$ 10$^{14}$ cm are remnants of the RT fingers due to the forward shock at 720 s.  Right panel:  velocities.  Some regions in the new RT instabilities reach velocities of $1.4 \times 10^9$ cm s$^{-1}$.}}
\end{figure*}

\subsection{Magnetar-Driven Instabilities}

We show densities and velocities 30 h after magnetar formation in Figure~\ref{fig:int}, after it has injected about 90\% of its spin-down energy into the bubble and its luminosity has fallen from $10^{48}$ to $10^{46}$ erg s$^{-1}$.  Because of the rapid release of nearly twenty times the energy of the original explosion, the bubble, which is at $1.25 \times 10^{14}$ cm, has almost overtaken the forward shock of the SN, which is at $1.6 \times 10^{14}$ cm.  \ken{Remnants of the RT fingers due to the forward shock at 720 s are now visible as the diffuse structures enclosed by the bubble at $r \sim 0.5 - 1.1 \times 10^{14}$ cm.  The fingers at $r \sim 1.3 - 1.6 \times 10^{14}$ cm are new RT instabilities due to the deceleration of the forward shock as it plows up the CSM.}  Their amplitudes and velocities have been strongly enhanced by pressure from the hot bubble behind them, which has acclererated iron-group elements in them to speeds of $1 - 2 \times 10^9$ cm s$^{-1}$, comparable to those of the forward shock.  This process could be the origin of the high-velocity \Ni\ and \Fe\ observed at unexpectedly early times in some CC SNe such as SN 1987A.  

\ken{Past simulations of SN 1987A by \citet{fryxell1991,kif06,candace2010,Ham10} showed  that significant mixing can occur during the explosion that partially explained its observed spectra but could not produce its high-velocity \Ni\ and \Fe\ lines.  More recent work by \citet{Mao15,Won15} produced high-velocity \Ni\ features but only under special conditions that depended on the choice of progenitor star and explosion energy.  \citet{Ono20} proposed a binary progenitor model and jet-like explosions for the origin of such lines that were quite different than the progenitor inferred for SN 1987A.}  

\ken{These models failed to reproduce high-velocity \Ni\ and \Fe\ lines because mixing in them is  mainly driven by the reverse shock before the forward shock breaks out the star.  The violence of the mixing depends on the structure of the progenitor and is greatest in red supergiants, whose extended envelopes allow instabilities to grow for longer times, dredging material deep in the ejecta up to higher velocities.  However, observations indicate that the progenitor of 1987A was probably a blue supergiant in which mixing is expected to be weak because the star is more compact.  Furthermore, the velocities observed for \Ni-rich ejecta are usually $\sim$ 4000 - 6000 km s$^{-1}$, which cannot be achieved by mixing due to the reverse shock (which decelerates rather than accelerates over time).  Energy from a magnetar or pulsar inside a CC SN can naturally explain the high-velocity \Ni\ and \Fe\ found in some explosions such as SN1987A.}

\begin{figure*}
\centering
\includegraphics[width=\columnwidth]{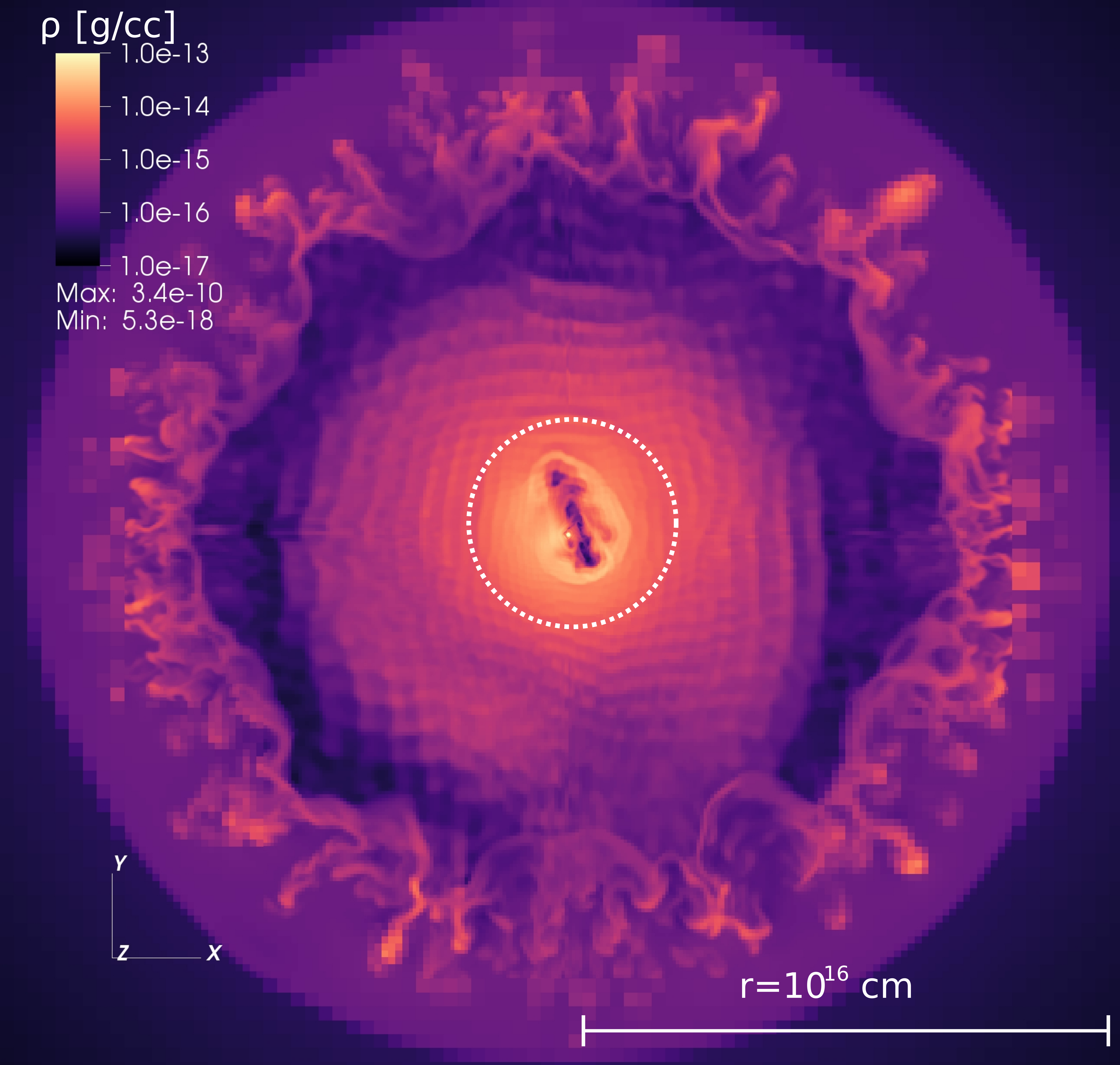} 
\includegraphics[width=\columnwidth]{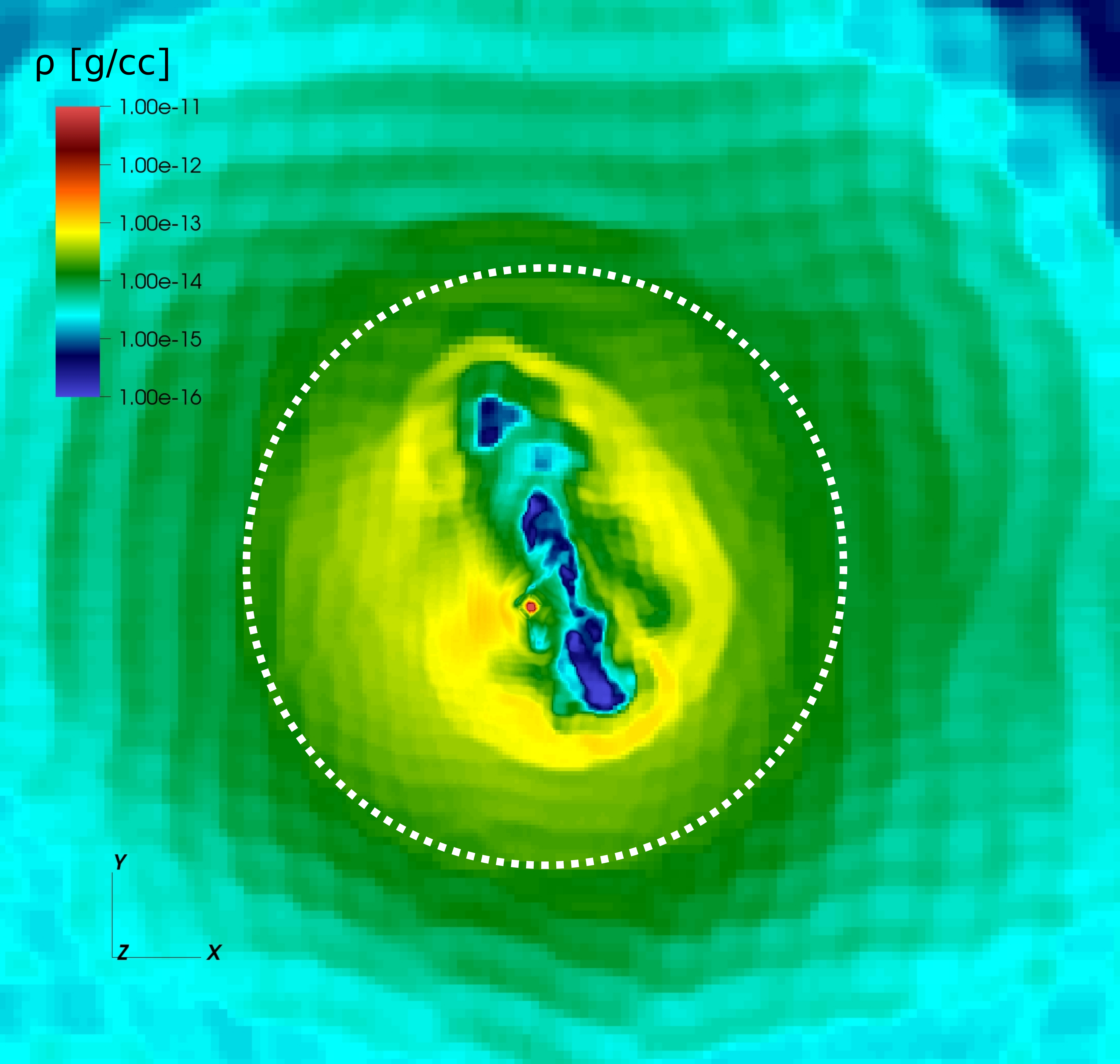} 
\caption{\label{fig:late}  200 days after magnetar birth.  Right panel:  mixing in the outer shell and magnetar bubble.  The inner most white dashed circle of radius $\sim$  $3 \times 10^{15}$ cm is the bubble, whose interior has become quite elongated.  Left panel:  close-up of the bubble.}
\end{figure*}

\begin{figure*}
\centering
\includegraphics[width=.8\textwidth]{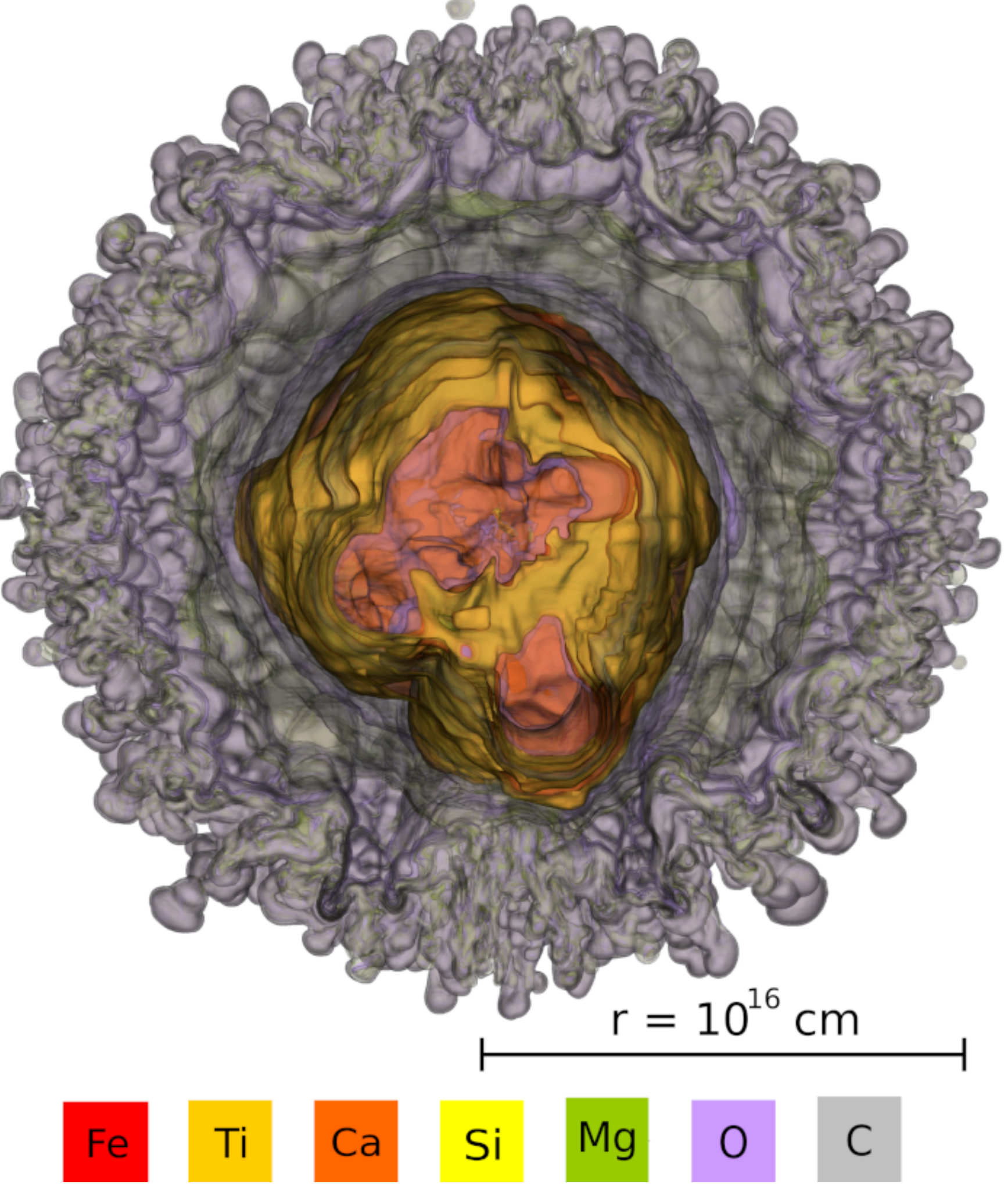} 
\caption{\label{fig:mixing} Distribution of elements in the ejecta at 200 days, each of which is represented by a color.  The isosurfaces of each element correspond to boundaries in mass fraction: Fe = 0.1, \Ca\ = 0.1, \Si\ = 0.05, \Ox\ = 0.3, \Cx\ = 0.03.  Here, Fe represents any remaining \Ni\ that was synthesized during the explosion and the \Co\ and \Fe\ into which the rest decayed.  The original \Cx, \Ox, and \Mg\ shells of the star have been mixed together in the outer regions while the innermost Fe, \Ti, \Ca, and \Si\ shells have been mixed together in the bubble.  Such mixing can greatly affect SN spectra.}
\end{figure*}

\begin{figure}
\centering
\includegraphics[width=\columnwidth]{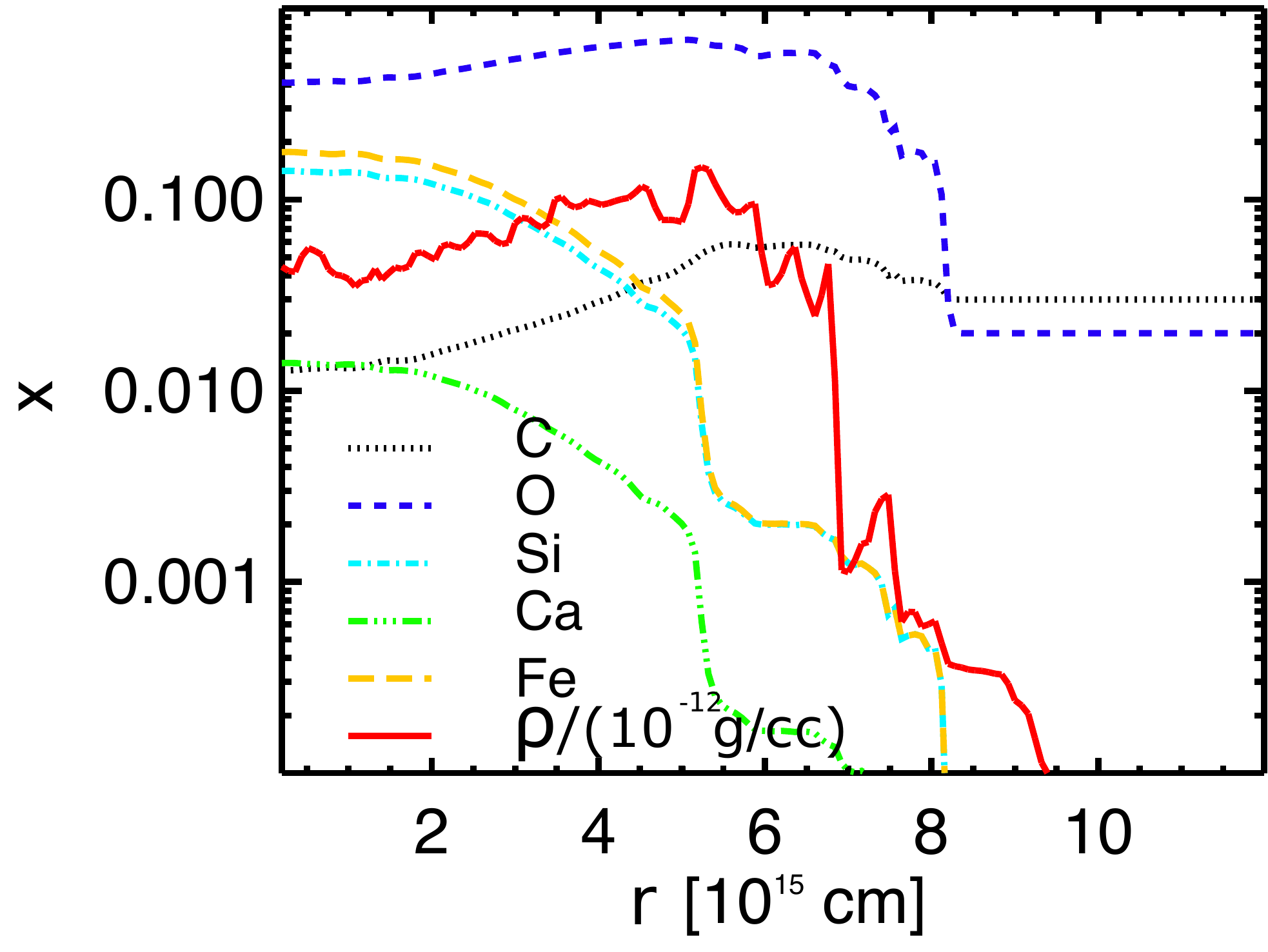} 
\includegraphics[width=\columnwidth]{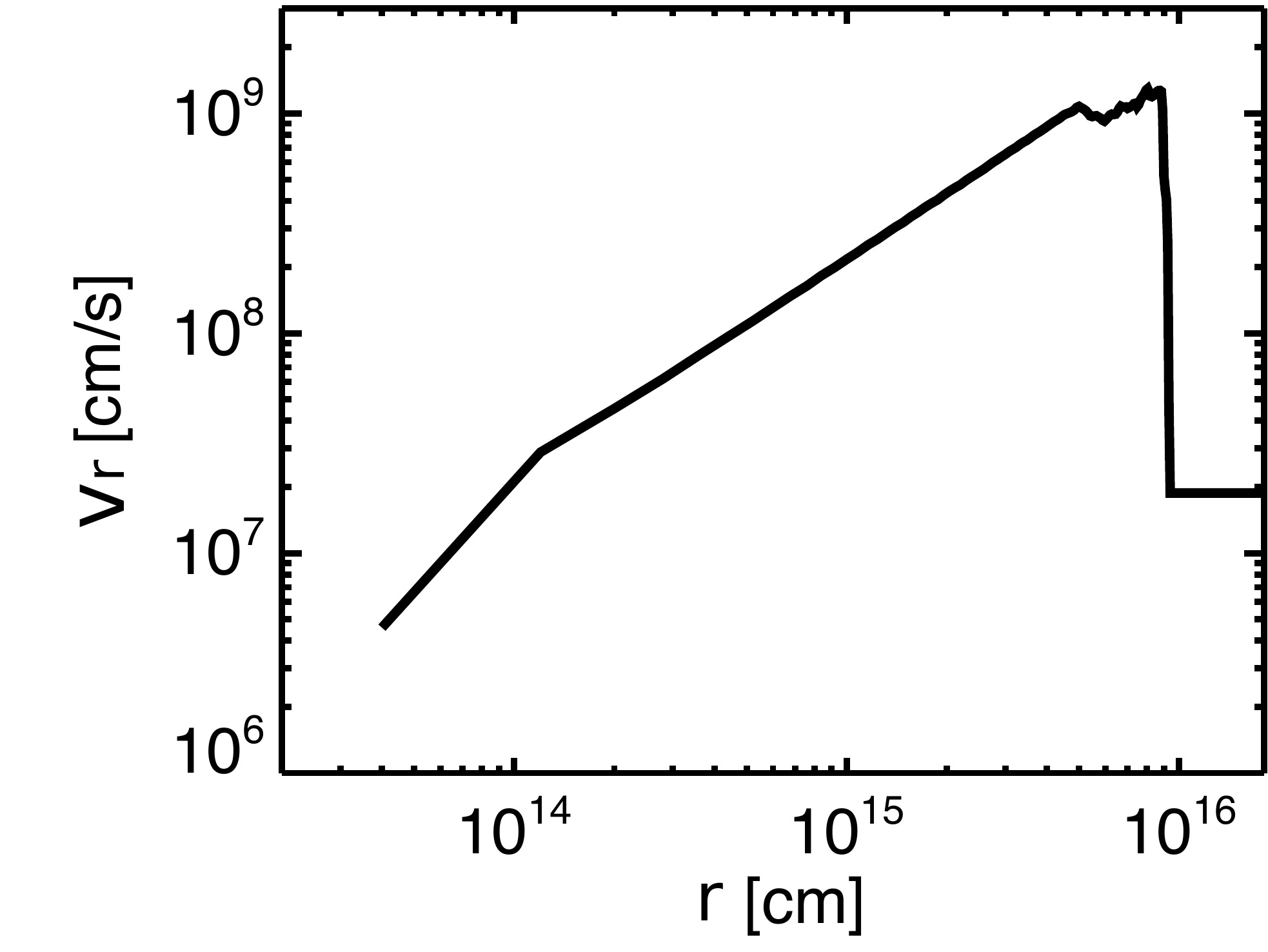} 
\caption{\label{fig:mixing_profile} 1D angle-averaged profiles of the SN ejecta at 200 days.  Upper panel:  densities and mass fractions. Some Fe has mixed with the \Cx\ and \Ox.  Lower panel:  velocities. The ripples at $6 - 10 \times 10^{15}$ cm are due to RT instabilities that are still evolving. The velocity profile reveals two free expansions, one deep within the magnetar bubble and one in the surrounding flow.}
\end{figure}

\subsection{Final Stages}

By 200 days the magnetar bubble and forward shock are at $\sim$  $3 \times 10^{15}$ cm and $10^{16}$ cm, respectively.  Approximately 99\% of the energy of the magnetar has been released and the ejecta is now expanding homologously.  As shown in Figure~\ref{fig:late}, RT instabilities appear on the largest scales just behind the forward shock while the interior of the magnetar bubble has become elongated.  This asymmetry was seeded by turbulence in the bubble at early times, when hot gas along some lines of sight preferentially expanded through channels of lower density in the ejecta.  \djw{A somewhat boxlike structure enclosing the wind bubble is visible (and more prominent in the right panel of Figure~\ref{fig:late}).  It arises because there is less AMR refinement in the low densities surrounding the bubble because of the relatively small density gradients there.  This structure becomes more distorted as the grid is doubled and is therefore ultimately an artifact of the mesh.}  The morphology of the magnetar bubble begins to freeze in mass coordinate at this time while the RT instabilities in the outer regions can continue to grow, depending on the CMS.  

We show how some of the elements have become mixed at this time in Figure~\ref{fig:mixing}.  The Fe represents any remaining \Ni\ that was synthesized during the explosion and the \Co\  and \Fe\ into which the rest decayed.  Most of the mixing has occurred between \Cx\ and \Ox\ in the outer regions and \Mg, \Si, \Ca\ and Fe in the bubble, but some Fe has been dredged up to $\sim 8 \times 10^{15}$ cm.  1D angle-averaged profiles of density, velocity, and elemental abundances are shown in Figure~\ref{fig:mixing_profile}. The flat mass fractions at small radii suggest that the inner ejecta is approaching chemical homogeneity. Fe at mass fractions of $10^{-3} - 10^{-2}$ has high velocities, $\sim 10^9$ cm s$^{-1}$.  The velocity profile reveals two free expansions, that of the central bubble (which continues to be pumped with energy at low levels by the magnetar) and that of the surrounding ejecta.  The ripples at the outer edge of the profile are due to the RT fingers.  The radius of the inner free expansion ($\sim$  $1.5 \times 10^{14}$ cm) is much smaller than that of the bubble because it is the gas most directly exposed to radiation from the now much dimmer magnetar. 
 
\subsection{\citet{Che16}}

\citet{Che16} modeled the same magnetar studied here except in 2D and only at early times in its evolution.  They omit the CSM of the star because they only evolved the explosion for short times after shock breakout.  They found that mixing in the shell of the magnetar bubble is more efficient in 2D, with many thin filamentary structures forming in contrast to the smoother clumpy structures found in 3D.  This primarily is due to differences in turbulent cascades in 2D and 3D.  

In the absence of a CSM, when radiation can escape the ejecta at early times, mixing in the magnetar bubble could determine the spectrum and light curve of the event because it sets the temperatures and structures of the clumps emitting high-energy photons from the explosion.  However, most of these events are expected to occur in a dense envelope so radiation breakout would happen much later.  As in our 3D models, iron-group elements deep in the ejecta were accelerated to high velocities at early times by energy from the magnetar in \citet{Che16}, reaching speeds of $1.2 \times 10^9$ cm s$^{-1}$. 

\section{Conclusion}

\dw{Our simulations clearly demonstrate that 3D simulations in full 4$\pi$ geometries are required to properly model the light curves and spectra of magnetar-powered SLSNe because both are required to reproduce hydrodynamical instabilities and to capture low-mode anisotropies, like those in the magnetar bubble at later times.}  We find that hydrodynamical instabilities arise on two scales from early times in the explosion:  in the forward shock of the original SN and in the much smaller bubble heated by the magnetar.     

Two results in particular stand out from our models.  First, energy from a magnetar could be the origin of the \Ni\ and \Fe\ observed at $\sim$ 4000 - 6000 km s$^{-1}$ in ordinary CC SNe such as SN 1987A, not just SLSNe.  Ejecting $0.2 - 0.4$ \Msun\ of \Ni\ at these velocities only requires $\sim 10^{49} - 10^{50}$ erg, which is small in comparison to typical CC SN explosion energies of $\sim 1.2 \times 10^{51}$ erg.  Even a weak magnetar with a spin period of 5-10 ms can provide this energy if most of the luminosity of the NS goes into accelerating the inner ejecta.  As discussed earlier, mixing due to the formation of a reverse shock by itself cannot account for the early appearance of iron-group elements observed in some CC SNe.  If this happens in magnetar-powered SLSN it would be manifested as broadened line features in \Fe\, \Ca, and \Si.  However, these elements may not be visible at early times if the explosion is shrouded by a dense CSM that traps radiation from the blast until later times.

Second, energy from the magnetar can drive much stronger mixing at intermediate times than in normal CC SNe because the hot magnetar bubble enhances the strength of RT instabilities about a day after the explosion.  This mixing can also dredge up heavier elements from greater depths so that they appear at earlier times when radiation does break out of the ejecta, even in the presence of a CSM.  Magnetar-powered mixing in 3D must be taken into account in producing accurate light curves and spectra for these events.

Indeed, strong mixing in magnetar-powered SNe could be the key to distinguishing them from other SLSN candidates such as pair-instability (PI) SNe \citep{barkat1967,heger2002,kasen2011,wet12a,wet12b,chatz2012,dessart2013}.  Mixing in the ejecta determines the times that specific elements appear in the spectrum of the explosion.  \citep{candace2010-2,chen2011,chen2014a} modeled PI SNe in 2D in \CASTRO.  They found that heating due to the decay of \Ni\ drives a shell at the boundary of the Ni--rich ejecta outward but this does not lead to much mixing, even in red supergiants with extended envelopes.  Iron-group elements therefore remain much deeper in mass coordinate in PI SN ejecta than in magnetar-powered SNe and only appear later, which will differentiate their spectra at intermediate and late times.  

\dw{We note that radiation from these events could be detected at a variety of times after the birth of the magnetar, ranging from just after shock breakout from the surface of the star to delayed breakout from an optically-thick envelope ejected by the star prior to explosion.  The key is the mass of the envelope and the time at which it is ejected.  If it has a low mass and is ejected at early enough times to become geometrically diluted by expansion then it may basically be transparent by the time the explosion occurs.  If it has a high mass and is ejected just before the explosion, radiation could be trapped for tens of days.  This will strongly affect the light curve of the SN and provide a diagnostic of its ambient environment.}

We adopted an idealized model for the magnetar in our study to focus on its effect on the dynamics of the ejecta.  Its luminosity was nearly isotropic, its magnetic field was constant, the dipole formula was used to represent its energy losses, and all this energy was emitted as radiation, not particles.  In reality, emission is likely anisotropic and the strength and orientation of the magnetic field probably evolves over time, with significant departures from the dipole approximation in energy losses.  Directionality in energy injection by the magnetar will lead to larger asymmetries in the flows than in our simulations, with important consequences for the observational signatures of the event. 

Furthermore, modeling the properties of the magnetar at birth remains extremely challenging.   
They are thought to form from rapidly rotating iron cores but there is no self-consistent treatment of rotation in any 1D stellar evolution code.  Although \citet{Mos15} recently attempted to follow the collapse of a rapidly rotating iron core with general-relativistic magnetohydrodynamic (GRMHD) simulations, they found it difficult to produce a magnetar without the use of an unrealistically large seed magnetic field.  Multidimensional stellar evolution models will be the next step to predicting initial conditions for magnetars from first principles.     

Although our simulations yield new insights into the hydrodynamics of magnetar-powered SLSNe, multidimensional radiation hydrodynamics will eventually be required to properly model shock breakout from the star (and CSM at later times) and how high-energy photons from the magnetar are thermalized in the flow, particularly in the magnetar bubble, which is filled with hot gas and is completely radiation dominated.  This physics will also be crucial to synthesizing realistic light curves and spectra for these events but it requires the electromagnetic spectrum of the magnetar itself as an input for which there is currently no self-consistent model.  Nevertheless, improved models and new observational data from SN factories such as the {\it Zwicky Transient Facility} and the {\it Vera C. Rubin Observatory} will soon reveal the central engines of SLSNe.

\acknowledgments

We thank Tug Sukhbold for providing the \KEPLER\ progenitor model and Dan Kasen, Ann Almgren, Lars Bildsten, Ken Nomoto, and Keiichi Maede for many useful discussions.  This research was supported by an EACOA Fellowship and by the Ministry of Science and Technology, Taiwan, R.O.C. under Grant no. MOST 107-2112-M-001-044-MY3. K.~C. thanks the hospitality of the Aspen Center for Physics, which is supported by NSF PHY-1066293, the Kavli Institute for Theoretical Physics, which is supported by NSF PHY-1748958, the Nordic Institute for Theoretical Physics (NORDITA), and Physics Division at the National Center for Theoretical Sciences (NCTS).  D.~J.~W. was supported by the Ida Pfeiffer Professorship at the Institute of Astrophysics at the University of Vienna.  Our numerical simulations were performed at the National Energy Research Scientific Computing Center (NERSC), a U.S. Department of Energy Office of Science User Facility operated under Contract No. DE-AC02-05CH11231, the Center for Computational Astrophysics (CfCA) at National Astronomical Observatory of Japan (NAOJ), and the TIARA Cluster at the Academia Sinica Institute of Astronomy and Astrophysics (ASIAA).

\software{\CASTRO\  \citep{Alm10, Zha11},} \KEPLER\ \citep{kepler, heger2010}

\end{document}